\documentclass[12pt]{article}
\usepackage{times}
\usepackage{geometry}
\usepackage[utf8]{inputenc}
\usepackage{enumitem,amssymb}
\usepackage{ragged2e}
\usepackage{pifont}
\usepackage{booktabs}
\usepackage{amsmath}
\usepackage{amssymb}
\usepackage{graphicx}
\usepackage{filecontents}
\usepackage[numbers]{natbib}

\def\farcs{\hbox{$.\!\!^{\prime\prime}$}}
\def\ltsim{\mathrel{\hbox{\rlap{\hbox{\lower3pt\hbox{$\sim$}}}\hbox{\raise2pt\hbox{$<$}}}}}
\def\gtsim{\mathrel{\hbox{\rlap{\hbox{\lower3pt\hbox{$\sim$}}}\hbox{\raise2pt\hbox{$>$}}}}}

\def\mAB{$m_{\rm AB}$}

\begin{document}

\begin{raggedright}

%

\LARGE
An Ultra Deep Field Survey with WFIRST\linebreak

\normalsize


%

\textbf{Authors:}
Anton M. Koekemoer (STScI),
R. J. Foley (UCSC),
D. N. Spergel (Princeton/CCA),
M. Bagley (UT Austin)
R. Bezanson (Pittsburgh),
F. B. Bianco (NYU),
R. Bouwens (Leiden),
L. Bradley (STScI),
G. Brammer (NBI),
P. Capak (Caltech),
I. Davidzon (Caltech),
G. De Rosa (STScI),
M. E. Dickinson (NOAO),
O. Dor{\'e} (JPL),
J. S. Dunlop (ROE),
R. S. Ellis (UCL),
X. Fan (Arizona),
G. G. Fazio (CfA),
H. C. Ferguson (STScI),
A. V. Filippenko (UCB),
S. Finkelstein (UT Austin),
B. Frye (Arizona),
E. Gawiser (Rutgers),
N. A. Grogin (STScI),
N. P. Hathi (STScI),
C. M. Hirata (OSU),
R. Hounsell (U. Penn),
G. D. Illingworth (UCSC),
R. A. Jansen (Arizona),
M. Jauzac (Durham),
S. W. Jha (Rutgers),
J. S. Kartaltepe (RIT),
A. G. Kim (LBL),
P. Kelly (Minnesota),
J. W. Kruk (NASA GSFC),
R. Larson (UT Austin),
J. Lotz (Gemini),
R. Lucas (STScI),
S. Malhotra (NASA GSFC),
K. Mandel (Cambridge),
R. Margutti (Northwestern),
D. Marrone (Arizona),
R. J. McLure (ROE),
K. McQuinn (UT Austin),
P. Melchior (Princeton),
B. Mobasher (UCR),
J. R. Mould (Swinburne),
L. Moustakas (JPL),
J. A. Newman (Pittsburgh),
C. Papovich (Texas A\&M),
M. S. Peeples (STScI/JHU),
S. Perlmutter (LBL),
N. Pirzkal (STScI),
J. Rhoads (NASA GSFC),
J. Rhodes (JPL),
B. Robertson (UCSC/IAS),
D. Rubin (STScI),
R. Ryan (STScI),
D. Scolnic (Duke),
A. Shapley (UCLA),
R. Somerville (Rutgers/CCA),
C. Steinhardt (NBI),
M. Stiavelli (STScI),
R. Street (LCO),
M. Trenti (Melbourne),
T. Treu (UCLA),
L. Wang (Texas A\&M),
Y. Wang (Caltech/IPAC),
D. Whalen (Portsmouth),
R. A. Windhorst (Arizona),
E. J. Wollack (NASA GSFC),
H. Yan (Missouri)
\linebreak

\end{raggedright}

\noindent
\textbf{Abstract:}

\noindent
Studying the formation and evolution of galaxies at the earliest cosmic times, and their role in reionization, requires the deepest imaging possible. Ultra-deep surveys like the HUDF and HFF have pushed to mag \mAB$\,\sim\,$30, revealing galaxies at the faint end of the LF to $z$$\,\sim\,$9$\,-\,$11 and constraining their role in reionization. However, a key limitation of these fields is their size, only a few arcminutes (less than a Mpc at these redshifts), too small to probe large-scale environments or clustering properties of these galaxies, crucial for advancing our understanding of reionization. Achieving HUDF-quality depth over areas $\sim$100 times larger becomes possible with a mission like the Wide Field Infrared Survey Telescope (WFIRST), a 2.4-m telescope with similar optical properties to HST, with a field of view of $\sim$1000 arcmin$^2$, $\sim$100$\times$ the area of the HST/ACS HUDF.
  
This whitepaper motivates an Ultra-Deep Field survey with WFIRST, covering $\sim$100$\,-\,$300$\times$ the area of the HUDF, or up to $\sim$1 deg$^2$, to \mAB$\,\sim\,$30, potentially revealing thousands of galaxies and AGN at the faint end of the LF, at or beyond $z$\,$\sim$\,9$\,-\,$10 in the epoch of reionization, and tracing their LSS environments, dramatically increasing the discovery potential at these redshifts.

~\\
\noindent
{\small \it Note: This paper is a somewhat expanded version of one that was submitted as input to the Astro2020 Decadal Survey, with this version including an Appendix (which exceeded the Astro2020 page limits), describing how the science drivers for a WFIRST Ultra Deep Field might map into a notional observing program, including the filters used and exposure times needed to achieve these depths.}

\pagebreak

\section{Introduction and Science Motivation}

Probing the formation and evolution of galaxies up to redshifts $\sim\,$9$\,-\,$10 and beyond, into the epoch of reionization, requires deep imaging to magnitudes \mAB$\,\gtsim\,$29.5$\,-\,$30 at optical to near-IR wavelengths, with spatial resolution of $\sim\,$0$\farcs$1 or better, to resolve the kpc-scale (or smaller) star-forming clumps at these redshifts. Achieving these depths is crucial for reaching sub-$L^{\star}$ galaxies (faint end of the galaxy luminosity function, hereafter `LF') at $z$\,$\sim\,$9$\,-\,$10 to explore their role in reionization and trace their assembly into more massive galaxies over cosmic time
  \cite{Bromm_2011ARA&A..49..373B,
	Madau_2014ARA&A..52..415M,
	Stark_2016ARA&A..54..761S}.
To date, the deepest fields have been obtained with HST, reaching these depths in the Hubble Ultra Deep Field (HUDF) initially in 2004 with ACS at optical wavelengths
  \cite{Beckwith_2006AJ....132.1729B,
	Oesch_2007ApJ...671.1212O},
subsequently in 2009$\,-\,$2012 with WFC3 at near-IR wavelengths
  \cite{Oesch_2010ApJ...709L..16O,
	Ellis_2013ApJ...763L...7E,
	Koekemoer_2013ApJS..209....3K,
	Illingworth_2013ApJS..209....6I}
and in 2013 at UV wavelengths
  \cite{Teplitz_2013AJ....146..159T,
	Rafelski_2015AJ....150...31R},
revealing a wealth of detail about the evolution of the faint end of the LF and size evolution of galaxies, as well as measuring the escape fraction of ionizing photons at high redshift and the role played by faint galaxies in reionization
  \cite{Oesch_2010ApJ...709L..21O,
	Bouwens_2010ApJ...709L.133B,
	Trenti_2010ApJ...714L.202T,
	Finkelstein_2012ApJ...756..164F,
	Finkelstein_2012ApJ...758...93F,
	Yan_2012ApJ...761..177Y,
	Oesch_2013ApJ...773...75O,
	Robertson_2013ApJ...768...71R,
	Dunlop_2013MNRAS.432.3520D,
	Ono_2013ApJ...777..155O,
	Bouwens_2014ApJ...793..115B,
	Tilvi_2014ApJ...794....5T,
	Finkelstein_2015ApJ...810...71F,
	Mei_2015ApJ...804..117M,
	Curtis-Lake_2016MNRAS.457..440C,
	Rutkowski_2016ApJ...819...81R,
	Bagley_2017ApJ...837...11B,
	Oesch_2018ApJ...855..105O}.
With the HUDF having essentially reached the maximum observational depth practically achievable for HST (totalling several Msec of observing time), the Hubble Frontier Fields program (HFF)
  \cite{Lotz_2017ApJ...837...97L}
was subsequently carried out, using gravitational lensing in six cluster fields to push up to an order of magnitude deeper into the LF at $z$\,$\sim$\,9$\,-\,$11, see also
  \cite{Oesch_2015ApJ...808..104O,
	Huang_2016ApJ...823L..14H,
	Livermore_2017ApJ...835..113L,
	Bouwens_2017ApJ...843..129B,
	Hoag_2018ApJ...854...39H},
although the relatively small lensing area results in much smaller volumes being probed at high redshift.

A key limitation of existing fields like the HUDF concerns its size, only $\sim\,$2$\,-\,$3 arcmin across, corresponding to $\sim\,$0.5$\,-\,$0.7 Mpc at $z$$\,\sim\,$10. This volume of space probed is too small to enable significant studies about the large-scale environment or clustering properties of galaxies at the faint end of the luminosity function, yet these questions are crucial to address in order to further advance our understanding of reionization. JWST will probe deeper (e.g.,
  \cite{Mason_2015ApJ...813...21M} and references therein),
likely over similar sized survey areas as those envisioned, for example, by current ERS and GTO programs, e.g.,
	`The NIRCam GTO Deep Field' (PI: M. Rieke),
	`The Cosmic Evolution Early Release Science (CEERS) Survey' (PI: S. Finkelstein),
	and
	`The JWST-NEP TDF' (PI: R. Windhorst,
  \cite{Jansen_2018PASP..130l4001J}).
Pure parallel programs with JWST could also potentially be carried out, along the lines of HST surveys such as BoRG
  \cite{Trenti_2011ApJ...727L..39T}
and WISP
  \cite{Atek_2010ApJ...723..104A};
by their nature such programs tend to be relatively shallow and are best suited for probing the sparse, bright end of the LF at high $z$. While larger surveys with HST have been carried out, these typically trade depth for area due to the constraints of observing time; for example the combined GOODS
  \cite{Giavalisco_2004ApJ...600L..93G}
and CANDELS
  \cite{Grogin_2011ApJS..197...35G,
	Koekemoer_2011ApJS..197...36K}
surveys cover $\sim$780 arcmin$^2$ to depths of \mAB$\,\sim\,$28.5 in up to ten bandpasses, while the COSMOS survey
  \cite{Scoville_2007ApJS..172...38S,
	Koekemoer_2007ApJS..172..196K}
covers $\sim\,$1.6$\,$deg$^2$ to \mAB$\,\sim\,$27.5 in a single bandpass (ACS F814W). These surveys have greatly enabled studies of galaxy formation at intermediate redshifts (eg, `cosmic noon' at $z$\,$\sim$\,2$\,-\,$3, as well as the brighter end of the LF up to $z$\,$\sim$\,6$\,-\,$8, with many of these results also reported in papers
  \cite{Oesch_2010ApJ...709L..21O}$\,-\,$\cite{Oesch_2018ApJ...855..105O}
and others), and with COSMOS furthermore enabling the first large-scale-structure dark matter maps obtained from weak lensing measurements
  \cite{Massey_2007Natur.445..286M}.
However, since these surveys are unable to probe down to the magnitude limits of $\sim$30 required to detect faint galaxies at $z$$\,\sim\,$9$\,-\,$10, the science is restricted to either the extremely bright end of the LF at these redshifts, or galaxy evolution at more moderate redshifts.
{\bfseries\slshape\boldmath To significantly advance knowledge of the faint end of the galaxy LF at reionization, it is necessary to obtain HUDF-quality imaging and depth (\mAB$\,\sim\,$30) over $\sim\,$100$\,-\,$300$\times$ larger area than existing ultra-deep surveys, achievable with a future facility like the Wide Field Infrared Survey Telescope (WFIRST).}

\section{Future Deep, Wide, High-Res. Imaging from Space: WFIRST}

\begin{figure}[t]
\centering
\includegraphics[width=0.65\textwidth]{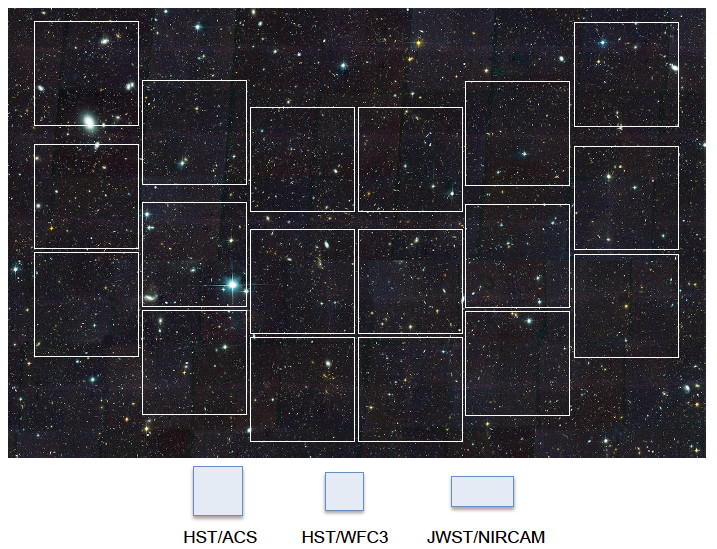}
\caption{\small Field of view of the WFIRST Wide Field Instrument (WFI), compared with instruments from HST and JWST. Each of the18 WFI detectors is a 4k$\times$4k HgCdTe array with 0$\farcs$11 per pixel.  The field of view of $\sim$1011$\,$arcmin$^2$ is about 100 times the area of the HST/ACS HUDF.}
\label{fig:wfi-fov}
\end{figure}

Achieving HUDF-quality depth over $\sim$100 times larger areas becomes possible with a mission like the Wide Field Infrared Survey Telescope (WFIRST), which is the top-ranked space mission from New Worlds, New Horizons 
  \cite{Spergel_2015arXiv150303757S,
	Akeson_2019arXiv190205569A}.
The design consists of a 2.4-m telescope, offering a comparable resolution to HST, equipped with the Wide Field Instrument (WFI) that is made up of 18 HgCdTe detectors, each 4k$\times$4k with 0$\farcs$11 pixels, providing a field of view over $\sim$1000$\,$arcmin$^2$, or almost 100 times the field of view of HST/ACS (Figure~\ref{fig:wfi-fov}). The currently planned configuration of the WFI includes seven broad-band filters ($RZY\!J\!H\!F$, and one ultra-wide) spanning $\sim$0.5$\,-\,$2.0~$\mu$m, an $R \approx 600$ grism (spanning 1.0$\,-\,$1.9\,$\mu$m), and an $R \approx 100$ prism (spanning 0.6$\,-\,$1.8\,$\mu$m), where the relative field of view is illustrated in Figure~\ref{fig:wfi-fov}.

\begin{figure}[t]
\centering
\includegraphics[width=0.7\textwidth]{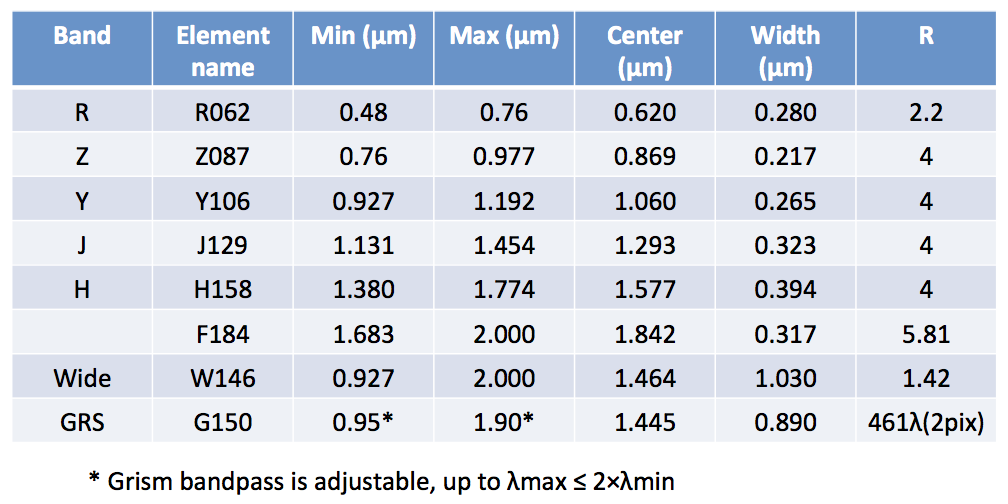}
\label{fig:wfi-filters}
\caption{\small Planned filter set for WFI, from the WFIRST Reference Information documents at
https://wfirst.gsfc.nasa.gov/science/WFIRST\_Reference\_Information.html}
\end{figure}

\section{Survey Programs with WFIRST}

The WFIRST design reference mission, further described in
  \cite{Spergel_2015arXiv150303757S,
	Akeson_2019arXiv190205569A},
includes a nominal 5-year observing plan containing a guest observing program, and several large surveys that target galactic science themes, namely exoplanets and bulge microlensing, as well as large extragalactic programs consisting of a high-latitude survey and a supernova program.

\subsection{The High Latitude Survey and Supernova Survey}

The high latitude survey (HLS) is envisioned to cover $\sim$2000$\,$deg$^2$ at sparse time-sampling cadence to depths of \mAB$\,\sim\,$26$\,-\,$27\,mag, and is thus at the opposite end of parameter space from that probed by ultra deep surveys, although it can be expected to yield significant numbers of rare high-luminosity sources in the epoch of reionization, along with measurements of large scale structure evolution on $\sim$\,Gpc-scales up to high redshift.

The supernova survey will cover $\sim\,$20$\,-\,$50$\,$deg$^2$ with more frequent time-sampling cadence, aiming to reach magnitudes \mAB$\,\sim\,$28$\,-\,$29 in the final full-depth images, achieving comparable depths to the medium-depth HST surveys such as GOODS and CANDELS, over $\sim\,$2$\,-\,$3 orders of magnitude larger area. This survey is subject to further optimization, but one current working version envisions two tiers: wide and deep, targeting different redshift ranges. If compatibility with ground-based facilities were not an issue, each tier would be split into a few fields, reducing the impact of cosmic variance, and all fields would be in the WFIRST Continuous Viewing Zone (CVZ). It is also possible that the deep-tier fields will be embedded in the wide-tier fields, reducing the impact of edge effects in the deep tier. Good choices for SNe fields (low extinction, low zodiacal emission, and CVZ) exist in both the North and the South.

\begin{figure}[t]
\centering
\includegraphics[width=0.85\textwidth]{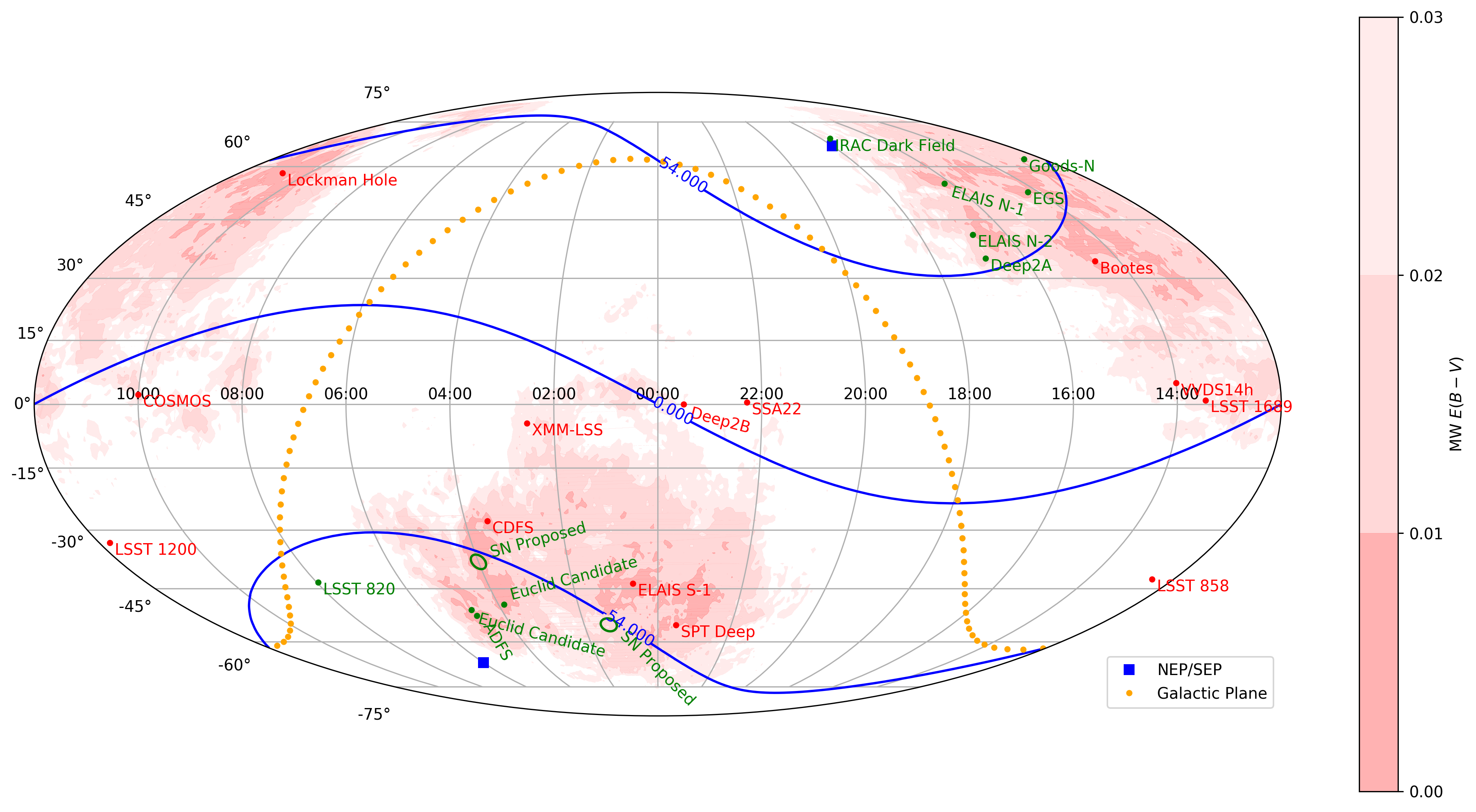}
\caption{Equatorial map of the sky indicating potential deep fields for WFIRST (courtesy of D. Rubin, R. Foley).  The red shading displays the Milky Way reddening (as indicated by the color bar on the right) with lower reddening values being darker.
Ecliptic latitudes of $|54^{\circ}|$, corresponding to the edge of the {\it WFIRST} continuous viewing zone (CVZ) are also displayed as blue lines.  Several extragalactic fields are marked, those in green being in the {\it WFIRST} CVZ,
while a representative set of other well-studied fields are in red, incl. CDFS, COSMOS, ELAIS-S1, SPT Deep, and other fields, all of which are collected in Table~\ref{tab:fields}.}
\label{fig:field-considerations}
\end{figure}

\subsection{GO Programs and Ultra Deep Science with WFIRST}\label{sec:GOfield}

While the HLS and SNe survey programs will lay significant groundwork for extragalactic galaxy and AGN astrophysics, the broader WFIRST Guest Observer (GO) community will likely seek to complement these fields with additional Ultra Deep Fields, covering a single or few pointings to fainter sensitivities. The WFIRST GO opportunity will arrive after both JWST and Euclid have been operational for several years, and therefore both the motivation and science return of such UDF programs should be carefully considered.

Scientifically, UDF fields with WFIRST will provide several critical capabilities. Contiguous, deep (\mAB$\,\sim\,$30 per filter per pointing) areas of the sky at 1$\,-\,$2\,$\mu$m with JWST will remain limited to a few hundred square arcminutes at most, even after years of operations, with deep JWST spectroscopy even more limited. Questions about the role of faint galaxies and AGN in cosmic reionization will depend on interpretations of disjoint fields with limited individual areas. While disjoint fields can help reduce cosmic variance for high-redshift samples, they cannot effectively connect the properties of galaxy populations to their surrounding environment.

A mission like WFIRST, providing contiguous, ultra-deep coverage over at least 1\,deg$^2$, will directly probe clustering and other spatial correlations for faint galaxies and AGN at significant comoving distances at high-$z$, along with possible variations in the faint-end slope of the LF with environment at high-$z$. This will likely provide the first clustering constraints on the dark matter halo mass for galaxies that dominate the luminosity density of the universe at early times. Single WFIRST pointings are wide enough to capture several ionized bubbles in the IGM during the height of the reionization era, and will supply a chance to connect the properties of the dominant ionizing sources with the ionization state of the IGM that surrounds them. Such studies combining WFIRST imaging and grism observations could connect the Ly$\alpha$ emission statistics of faint galaxy populations directly with their environmental overdensity. While the WFIRST HLS and SNe Deep Fields will also prove tremendously fruitful for studies of the reionization era and throughout the high-redshift universe, the ability to provide sensitive imaging over a substantial area through GO UDFs will remain a unique WFIRST role for at least a decade after HST and JWST.

\section{Potential Locations for Deep Fields with WFIRST}

For any potential deep field programs, important considerations include the fraction of time that the field is accessible to WFIRST (i.e., location relative to the CVZ), as well as accessibility to ancillary telescopes that will be expected to play significant roles in observing the field (ie, whether the fields are exclusively accessible to ground-based telescopes in the northern or southern hemispheres, or both). In addition, the amount of galactic extinction $E(B-V)$ is an important consideration, in order to avoid introducing reddening-related effects in colour measurements of supernovae and galaxies. The amount of zodiacal emission is also important to consider in terms of the ultimate limiting magnitude that can be reached in a given location.
The presence of deep X-ray, Herschel far-IR, and ALMA observations will also prove valuable given their unique probes of galaxy and AGN SEDs well outside the wavelength regime covered by WFIRST. Finally, connecting with 21cm surveys and submillimeter surveys, the HERA and SPT Deep fields are of particular interest for the placement of GO grism fields that could connect with ground-based intensity mapping experiments that directly probe the epoch of reionization in complementary ways.

\begin{table}[h!]
\caption{A Compilation of representative well-studied extragalactic fields}

\begin{tabular}{lrrrrlll}
\hline
\hline
Field 			& R.A.		& Dec. 		& Ecl. Lat. & Area (deg$^2$)	& E(B-V) & Rel. Zodi & Days/yr	\\
\hline
\hline
{\bf Polar fields ($<36^\circ$):}    \\
IRAC Dark Field		& 17:40		& +69:00	& +87	& 0.2	& 0.043		& 1.0	& 365	\\
Extended Groth Strip	& 14:17		& +52:30	& +60	& 0.2	& 0.009		& 1.2	& 365	\\
GOODS-N			& 12:36		& +62:13	& +57	& 0.25	& 0.012		& 1.2	& 365	\\
Deep2A			& 16:52		& +34:55	& +57	& 1 	& 0.018		& 1.2	& 365	\\
ELAIS N-2		& 16:46		& +41:01	& +63	& 5 	& 0.014		& 1.1	& 365	\\
ELAIS N-1		& 16:11		& +55:00	& +73	& 9 	& 0.008		& 1.0	& 365	\\
Akari Deep Field South	& 04:44 	& $-$52:20	& $-$73	& 12	& 0.008		& 1.0	& 365	\\
JWST-NEP-TDF		& 17:22		& +65:49	& +86	& 0.2	& 0.042		& 1.0	& 365	\\
NEP-Spitzer		& 18:00		& +66:33	& +90	& 10	& 0.046		& 1.0	& 365	\\
SEP-Spitzer		& 06:00 	& $-$66:33	& $-$90	& 10	& 0.062		& 1.0	& 365	\\
\hline
{\bf Equatorial fields}:    \\
CDFS			& 03:32 	& $-$27:48	& $-$45	& 0.3	& 0.008		& 1.4	& 229	\\
Deep2B			& 23:30		& +00:00	& +3	& 1 	& 0.044		& 19	& 146	\\
SSA22			& 22:17		& +00:24	& +10	& 4 	& 0.056		& 5.6	& 149	\\
COSMOS			& 10:00		& +02:12	& $-$9	& 2 	& 0.018		& 6.0	& 148	\\
VVDS14h			& 14:00		& +05:00	& +16	& 4 	& 0.026		& 3.6	& 153	\\
ELAIS S-1		& 00:35		& $-$43:40	& $-$43	& 7 	& 0.008		& 1.5	& 215	\\
Bootes			& 14:32		& +34:16	& +46	& 9 	& 0.016		& 1.4	& 236	\\
Lockman Hole		& 10:45		& +58:00	& +45	& 11	& 0.011		& 1.4	& 229	\\
XMM-LSS			& 02:31 	& $-$04:30	& $-$18	& 11	& 0.024		& 3.2	& 155	\\
SPT Deep		& 23:30		& $-$55:00	& $-$46	& 100	& 0.010		& 1.4	& 236	\\
HERA			& 07:00 	& $-$30:43	&	& 1200	&		&	&	\\
\hline
\end{tabular}
\label{tab:fields}
\end{table}

\section{Summary}

In summary, this paper presents a recommendation for an Ultra Deep Field program with WFIRST, representing a major leap forward from existing ultra-deep surveys by increasing the area by a factor of $\sim\,$100$\,-\,$300$\times$, covering $\sim$1\,deg$^2$ to \mAB$\,\sim$\,30 in six broad-band filters across 0.5$\,-\,$2$\,\mu$m, to study the impact and evolution of the faint galaxy population in the epoch of reionization to $z$$\,\sim\,$9$\,-\,$10 or beyond,  dramatically increasing the discovery potential at these redshifts.

\pagebreak

\section*{Appendix: Mapping Science Drivers to Observational Design}

Current observational constraints on galaxy populations at $z\sim\,$9$\,-\,$10 are primarily limited to the bright end of the UV LF, with detections generally brighter than \mAB$\,\sim\,$26.2$\,-\,$26.7 in the F160W filter, across search areas $\sim\,$500$\,-\,$700$\,$arcmin$^2$, with $\sim\,$10$\,-\,$15 candidates known to date
(e.g.,
  \cite{Bouwens_2016ApJ...830...67B,
	Oesch_2018ApJ...855..105O,
	Yung_2019MNRAS.483.2983Y} and references therein).
It is necessary to probe at least $\gtsim\,$3 magnitudes deeper in order to reach fainter galaxies more typical of the bulk of the galaxy population, as well as probing wider to cover sufficiently large volumes representative of LSS variations and ionization bubbles at these epochs. A key limitation in these previous surveys has been the relatively limited depth at wavelengths blueward of Ly$\alpha$, given that typical selection criteria include a requirement of non-detections in blueward bands that are $\gtsim\,$1.2$\,-\,$1.5 mag deeper than the detection bands.

In Table~\ref{tab:depths} we present an example of the depths that could be achieved with the WFIRST WFI in the six filters that cover its wavelength range from $\sim\,$0.5$\,-\,$2$\mu$m, with calculations based on the depths reported in
  \cite{Akeson_2019arXiv190205569A,
	Hounsell_2018ApJ...867...23H}
together with the current WFIRST Exposure Time Calculator
\footnote{http://www.stsci.edu/scientific-community/wide-field-infrared-survey-telescope/science-planning-toolbox/pandeia}.
It can be seen that in a reasonable amount of observing time, $\sim$350$\,$hours (comparable to a number of previous large HST programs), limiting depths of \mAB$\,\sim\,$31.4 (2$\sigma$) can be achieved in the filters blueward of the Lyman break that would provide the non-detections needed to confirm high-$z$ candidates, while detection limits \mAB$\,\sim\,$29.7$\,-\,$30 (5$\sigma$) are achievable in the two reddest filters, reaching $\gtsim\,$3$\,-\,$3.5 magnitudes deeper than the current samples of $z\sim\,$9$\,-\,$10 galaxies.

\begin{table}[h!]
\begin{center}
\caption{Example WFIRST UDF: limiting depths (\mAB)}

\begin{tabular}{ccll}
\hline
\hline
Filter 	& Exp. Time	& 5$\sigma$ limit$^a$	& 2$\sigma$ limit$^a$	\\
\hline
\hline
R062	& 30$\,$h	& 30.3			& 31.3			\\
Z087	& 60$\,$h	& 30.4			& 31.4			\\
Y106	& 70$\,$h	& 30.4			& 31.4			\\
J129	& 90$\,$h	& 30.4			& 31.4			\\
H158	& 40$\,$h	& 30.0			& 			\\
F184	& 60$\,$h	& 29.7			& 			\\
\hline
\multicolumn{4}{p{.45\textwidth}}{\scriptsize $^a$These values are preliminary and subject to change, pending further information on the specifications and performance of the WFIRST instrumentation, incl. detector and filter characteristics.}
\end{tabular}
\label{tab:depths}
%
\end{center}
\end{table}

\begin{figure}[h!]
\centering
\includegraphics[width=0.82\textwidth]{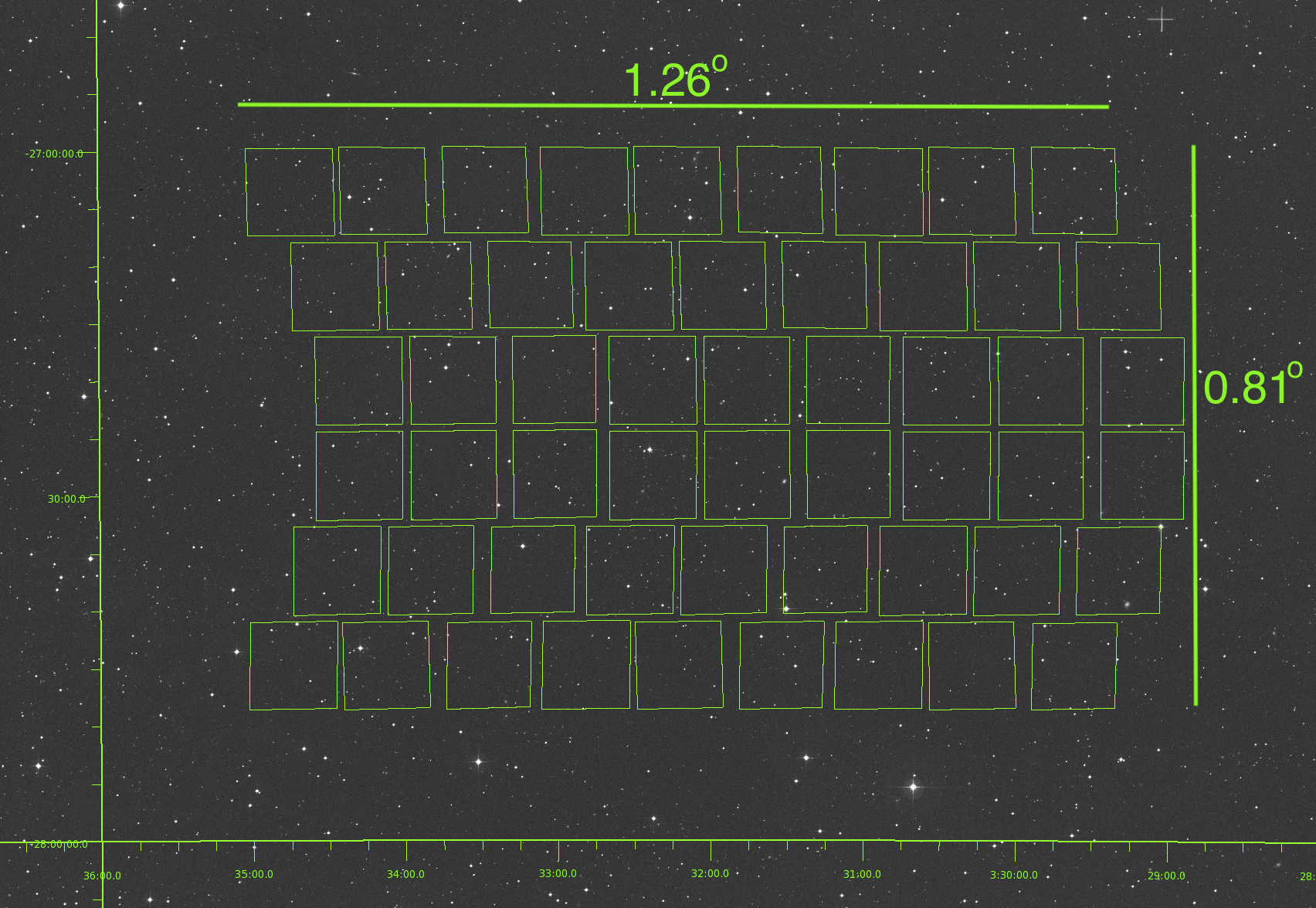}
\caption{\small Notional layout example for a $\sim$1$\,$deg$^2$ Ultra Deep Field with WFIRST/WFI, repeating the WFI footprint (3$\times$6 detectors, Fig.~\ref{fig:wfi-fov}, rotated 90$^\circ$) horizontally in a 3-point mosaic pattern, sampling $\sim\,$1.26$^\circ\times$0.81$^\circ$ (in the CDFS/HUDF as an example, but could be elsewhere). In this notional program, each of the 3 WFI pointings would be observed for 350$\,$hours in 6 filters (Table~\ref{tab:depths}), totalling 1050$\,$hours of exposure time. This would achieve 2$\sigma$ detection limits \mAB$\,\sim\,$31.4 in filters blueward of the Lyman break at z$\,\sim\,$9$\,-\,$10, with 5$\sigma$ detection limits as faint as \mAB$\,\sim\,$30 in the redward filters, enabling crucial two-band detection of high-$z$ dropout candidates at least $\gtsim\,$3$\,-\,$3.5 mag fainter than current samples in large surveys.}
\label{fig:wfi-udf}

\centering
\includegraphics[width=0.6\textwidth]{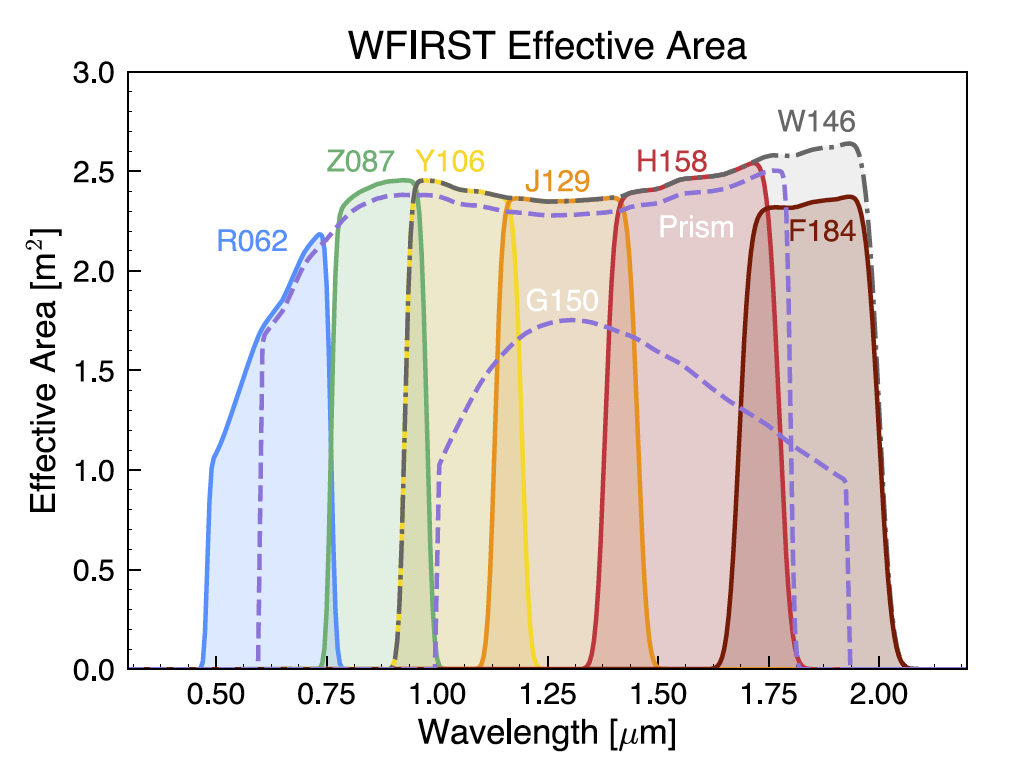}
\caption{\small Effective area and wavelength coverage of the relevant WFI filters 
\cite{Akeson_2019arXiv190205569A};
for comparison, the effective area for HST ACS is $\sim\,$1.9$\,$m$^2$ at optical wavelengths, and $\sim\,$2.4 m$^2$ for HST WFC3/IR.}
\label{fig:wfi-filter-curves}
\end{figure}

\noindent
Moreover, these depths are achieved over a single field of view of WFIRST/WFI (1011~arcmin$^2$). The coverage could be extended by obtaining, for example, 3 pointings (each totalling 350$\,$hours), either contiguously over an area extending $\sim\,$1.26$^\circ$$\times$0.81$^\circ$ (see Figure~\ref{fig:wfi-udf}) yielding a single $\sim\,$1$\,$deg$^2$ field, i.e. $\sim\,$300$\times$ the area of the current HUDF, or alternatively in multiple locations across the sky, e.g. with 3 separate fields each  $\sim\,$100$\times$ the area of the current HUDF. In either case,the total amount of exposure time would be 1050$\,$hours, not significantly above the largest HST observing programs to date (and $\sim$2.4\% of the nominal 5-year timespan currently planned for WFIRST).

\pagebreak

%

%

\bibliographystyle{unsrt}

\end{document}